\documentclass[a4paper,superscriptaddress,citeautoscript,twocolumn,prl,nobibnotes]{revtex4-1}

\usepackage[T1]{fontenc}
\usepackage{times}
\usepackage{mathptmx}
\usepackage{amsmath,amssymb}
\usepackage{graphicx,color}
\usepackage[a4paper,body={18cm,24.5cm},top=2cm]{geometry}

\begin{document}

\title{Asymmetric balance in symmetry breaking}

\author{Bruno Garbin}
\altaffiliation{now with Universit\'e Paris-Saclay, CNRS, Centre de Nanosciences et de Nanotechnologies, 91120 Palaiseau, France}
\affiliation{The Dodd-Walls Centre for Photonic and Quantum Technologies, New Zealand}
\affiliation{Physics Department, The University of Auckland, Private Bag 92019, Auckland 1142, New Zealand}
\author{Julien Fatome}
\affiliation{Laboratoire Interdisciplinaire Carnot de Bourgogne (ICB), UMR 6303 CNRS, Universit\'e de Bourgogne
Franche-Comt\'e, 9 Avenue Alain Savary, BP 47870, F-21078 Dijon, France}
\author{Gian-Luca Oppo}
\affiliation{SUPA and Department of Physics, University of Strathclyde, Glasgow G4 0NG, Scotland, EU}
\author{Miro Erkintalo}
\author{Stuart G. Murdoch}
\author{St\'ephane Coen}
\email{s.coen@auckland.ac.nz}
\altaffiliation{corresponding author}
\affiliation{The Dodd-Walls Centre for Photonic and Quantum Technologies, New Zealand}
\affiliation{Physics Department, The University of Auckland, Private Bag 92019, Auckland 1142, New Zealand}

\begin{abstract}
  \noindent Spontaneous symmetry breaking is central to our understanding of physics and explains many natural
  phenomena, from cosmic scales to subatomic particles. Its use for applications requires devices with a high level
  of symmetry, but engineered systems are always imperfect. Surprisingly, the impact of such imperfections has
  barely been studied, and restricted to a single asymmetry. Here, we experimentally study spontaneous symmetry
  breaking with two controllable asymmetries. We remarkably find that features typical of spontaneous symmetry
  breaking, while destroyed by one asymmetry, can be restored by introducing a second asymmetry. In essence,
  asymmetries are found to balance each other. Our study illustrates aspects of the universal unfolding of the
  pitchfork bifurcation, and provides new insights into a key fundamental process. It also has practical
  implications, showing that asymmetry can be exploited as an additional degree of freedom. In particular, it would
  enable sensors based on symmetry breaking or exceptional points to reach divergent sensitivity even in presence
  of imperfections. Our experimental implementation built around an optical fiber ring additionally constitutes the
  first observation of the polarization symmetry breaking of passive driven nonlinear resonators.
\end{abstract}

\maketitle

\section*{Introduction}

\noindent Spontaneous symmetry breaking (SSB) is a concept of fundamental importance \cite{anderson_more_1972,
strocchi_symmetry_2008, malomed_spontaneous_2013}. It is central to the standard model of particle physics
\cite{nambu_nobel_2009, englert_broken_1964, *higgs_broken_1964, bernstein_spontaneous_1974}, underpins phenomena as
diverse as ferromagnetism and superconductity \cite{landau_statistical_1980, tilley_superfluidity_1990,
leggett_concept_1991}, and plays a key role in convection cells and fluid mechanics \cite{crawford_symmetry_1991},
morphogenesis \cite{turing_chemical_1952}, embryo development \cite{korotkevich_apical_2017}, and more generally
self-organization \cite{prigogine_symmetry_1969}. SSB can also be exploited for many applications
\cite{kaplan_enhancement_1981, liu_spontaneous_2014, hamel_spontaneous_2015, cao_experimental_2017,
del_bino_symmetry_2017, woodley_universal_2018}. In particular, new unique ways to manipulate light have recently
been demonstrated in structures engineered to exhibit parity-time ($\mathcal{PT}$) symmetry breaking
\cite{feng_single-mode_2014, *hodaei_parity-timesymmetric_2014}. Engineered systems however often exhibit deviations
from perfect symmetry because of naturally occurring imperfections \cite{kondepudi_observation_1986,
dangelo_spatiotemporal_1992}. Surprisingly, the impact of asymmetries on SSB-related dynamics has barely been
considered in experiments, and essentially restricted to situations with only one asymmetry parameter
\cite{dangelo_spatiotemporal_1992, gelens_multistable_2010}. One exception are results obtained by Benjamin four
decades ago on Couette flow between rotating cylinders, which are clearly linked to the presence of two
imperfections, albeit only one was controlled \cite{benjamin_bifurcation_1978, *benjamin_bifurcation_1978-1}. Here,
we report for the first time an experimental study of a system that exhibits spontaneous mirror-symmetry breaking
with two fully controllable asymmetry parameters. While the characteristic dynamics of SSB --- random, spontaneous
selection between two mirror-like states with opposite symmetries --- are destroyed by one asymmetry, we remarkably
observe that a second asymmetry can restore them. In essence, the two asymmetries can balance each other.
Interestingly, this is conceptually related to the design principle of asymmetric balance, by which a design or an
art composition can be asymmetric, and yet, still look balanced \cite{pipes_foundations_2003}.

Spontaneous symmetry breaking is underlain by the fundamental pitchfork bifurcation \cite{strogatz_nonlinear_1994}.
For a system with left/right or mirror symmetry, that bifurcation describes how a symmetric state transitions to two
equivalent, stable, mirror-like asymmetric states [see, e.g., panel~(b) of Fig.~\ref{fig:principle}]. The pitchfork
is however a structurally fragile, degenerate bifurcation: in the presence of small asymmetries, one of the
asymmetric states dominates while the other cannot be spontaneously excited \cite{hirsch_invariant_1977,
benjamin_bifurcation_1978, *benjamin_bifurcation_1978-1, golubitsky_analysis_1979, *golubitsky_singularities_1985}.
It turns out that only \emph{two} parameters are needed to describe all the possible topologies of the perturbed
pitchfork --- its so-called universal unfolding \cite{benjamin_bifurcation_1978, *benjamin_bifurcation_1978-1,
golubitsky_analysis_1979, *golubitsky_singularities_1985, ball_understanding_2001, alam_universal_2004}. This
argument has been used to reduce the number of parameters in the search of simplified models of complex problems,
such as limb coordination \cite{park_imperfect_2008} --- a feature found in movements of a huge range of animals ---
or the emergence of the ubiquitous homochirality of biological molecules \cite{ball_life_2016,
*lebreton_molecular_2018}. It has also guided recent engineering research in buckling-resistant structures and led to
the discovery that optimal designs with imperfect symmetry only emerge when considering two asymmetry parameters
\cite{varkonyi_imperfect_2007}. We note that evolution, which is inherently guided by optimization, has produced
countless designs with near but not perfect symmetry, including the functional neural wiring of the brain
\cite{varkonyi_emergence_2006}. Clearly, considering two asymmetry parameters instead of one in the study of SSB can
have dramatic and intriguing consequences. To the best of our knowledge, our work is the first to experimentally
address how two asymmetries can balance each other.

\section*{Polarization symmetry breaking}

\begin{figure}[t]
  \centerline{\includegraphics{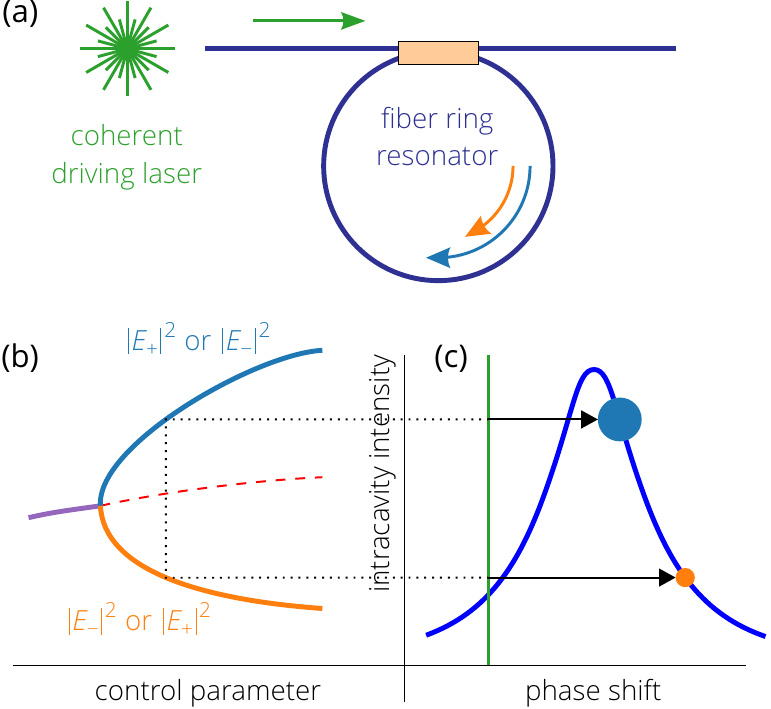}}
  \caption{Illustration of polarization symmetry breaking. (a) Schematic diagram of a driven passive
    nonlinear fiber ring resonator. (b) Pitchfork bifurcation diagram showing how the intensities of the two
    polarization modes part above a certain threshold. The control parameter can be either the driving power or the
    driving laser frequency. (c) Resonance of the system illustrating how in an asymmetric state the stronger
    mode (large blue dot) can undergo a smaller effective nonlinear shift (black arrow) and be closer to resonance
    than the weaker mode (small orange dot). This occurs when the driving laser (green line) is red-detuned (left
    side of the resonance).}
  \label{fig:principle}
\end{figure}

\noindent Our experiment is based on a passive nonlinear optical fiber ring resonator (akin to a Fabry-P\'erot
etalon) that presents two distinct orthogonal polarization modes. The resonator is externally, coherently driven by
intense laser light [Fig.~\ref{fig:principle}(a)] so as to excite both of these modes; hereafter $E_+$ and~$E_-$
denote the modes' complex electric field amplitudes inside the resonator. Ideally, when the two polarization modes
are equally driven and are degenerate (i.e., the resonator material is isotropic, and the modes have identical
resonance frequencies), the system is symmetric with respect to an interchange of the two modes, $E_+
\rightleftarrows E_-$. In the simplest case, the stationary intracavity field solution assumes that symmetry, $E_+ =
E_-$, and the two modes have the same intensities. Symmetry breaking occurs above a certain threshold
\cite{areshev_polarization_1983, haelterman_polarization_1994}, and manifests itself by the parting of the
intensities of the two polarization modes, $|E_+|^2 \neq |E_-|^2$ [Fig.~\ref{fig:principle}(b)]: the symmetric
solution loses its stability in favor of two mirror-like asymmetric solutions. The instability stems from the cubic
(Kerr) nonlinearity of silica optical fibers \cite{agrawal_nonlinear_2013}, by which the phase of one mode is
affected by the intensity of the other (cross-phase modulation, or XPM). Critically, in optical fibers, XPM between
polarization modes can be stronger than self-phase modulation (SPM) so that the weaker mode can experience a larger
nonlinear phase shift than the stronger mode \cite{agrawal_nonlinear_2013, gilles_polarization_2017}. In these
conditions, the weaker mode is pushed away from resonance while the stronger mode is pulled towards it, reinforcing
any initial intensity imbalance [Fig.~\ref{fig:principle}(c)] \cite{del_bino_symmetry_2017}. This polarization SSB is
formally identical to the SSB that occurs in the same system when considering two counter-propagating beams
\cite{kaplan_directionally_1982}, and which was recently studied experimentally \cite{del_bino_symmetry_2017}. In
both cases, an imbalance of driving power between the two modes (beams) readily provides a controllable asymmetry
parameter. In our experiment, we have also manipulated the wavenumbers of the two driven polarization components as
the second asymmetry parameter.

Passive driven scalar Kerr resonators can be efficiently described by a mean-field
approach~\cite{lugiato_spatial_1987, haelterman_polarization_1994}. For two incoherently coupled polarization modes,
assuming continuous-wave (cw) driving and neglecting chromatic dispersion, the evolution of $E_+$ and $E_-$ over
time~$t$ is given by (using the same normalization as in~\cite{leo_temporal_2010}),
\begin{align}
  \frac{\partial E_+}{\partial t} = \left[ -1 + i(|E_+|^2 + B|E_-|^2 - \Delta_+)\right] E_+ + \sqrt{X} \cos\chi\,,
  \label{eq:LL1}\\
  \frac{\partial E_-}{\partial t} = \left[ -1 + i(|E_-|^2 + B|E_+|^2 - \Delta_-)\right] E_- + \sqrt{X} \sin\chi\,.
  \label{eq:LL2}
\end{align}
In these equations, the incoherent coupling between the two modes is determined by the XPM coefficient~$B$ ($B$ > 1).
Other terms on the right-hand side represent, respectively, losses, SPM, the detuning of the driving frequency with
respect to resonance, and the driving strengths of each mode. Here $X$ represents the total driving power, while a
driving power imbalance between the modes is accounted for by introducing an effective driving polarization
ellipticity angle~$\chi$. An ellipticity angle $\chi$ of $45^\circ$ represents perfectly balanced driving. Because of
residual birefringence in our fiber resonator, the resonances of the two polarization mode families are normally
observed for different driving laser frequencies, which correspond to having different detunings in the equations
above, $\Delta_+ \neq \Delta_-$. The difference in detuning, $\delta\Delta = \Delta_+ - \Delta_-$, equivalently
represents the difference in wavenumbers with which the two polarization components propagate inside the resonator
(see Appendix~A) and is our second asymmetry parameter. It is tuned in our experiment by shifting the carrier
frequency of the $E_-$ mode with a frequency shifter as explained below. Symmetry for the interchange of the two
modes in Eqs.~(\ref{eq:LL1})--(\ref{eq:LL2}) is obtained with $\chi=45^\circ$ and $\delta\Delta=0$. Note also the
absence of rotational phase invariance in these equations because of the external driving terms, which is a key
feature of the passive driven resonator considered here, in contrast to, e.g., laser resonators.

\section*{Experimental setup}

\begin{figure*}
  \centerline{\includegraphics{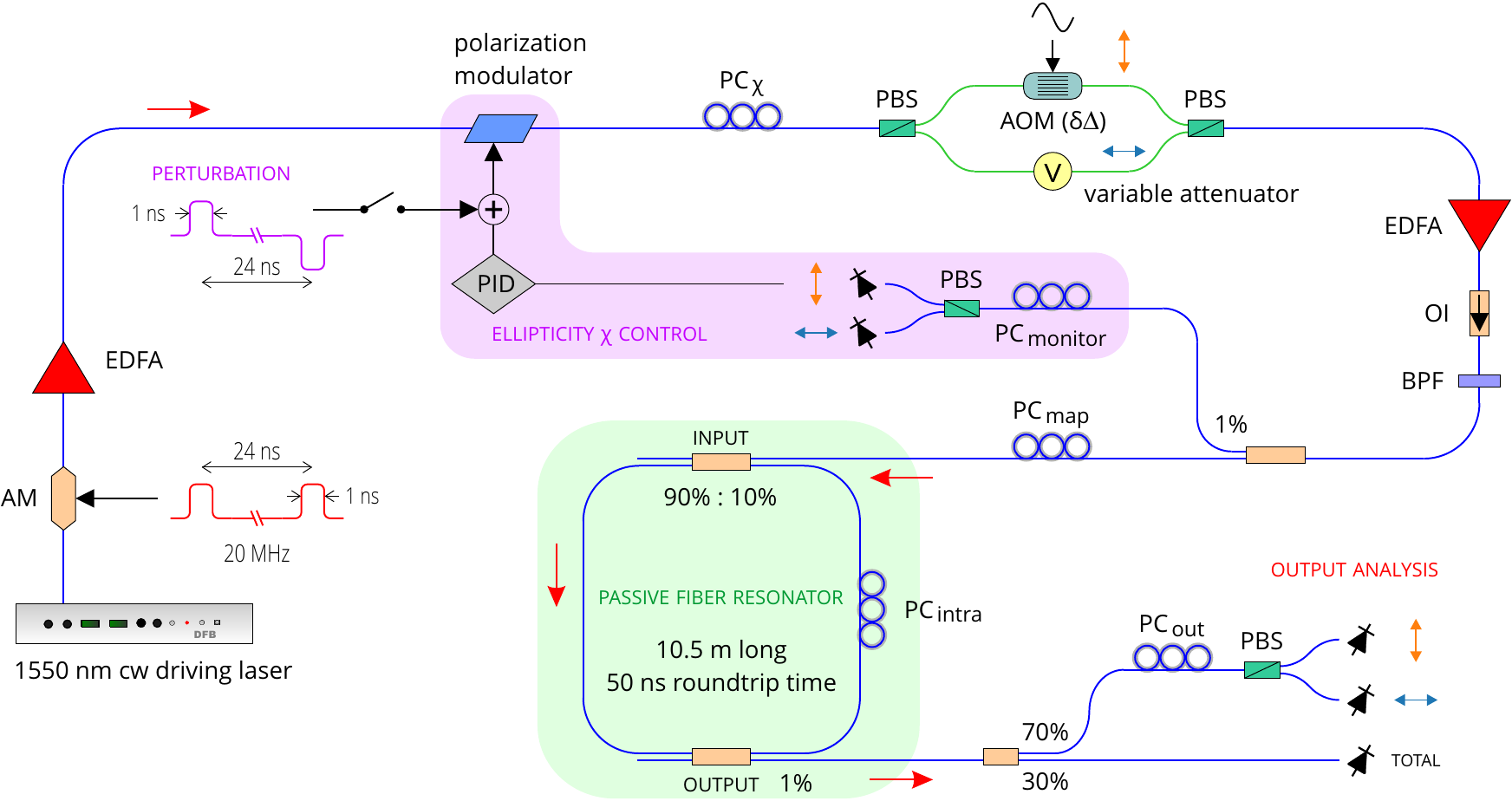}}
  \caption{Experimental setup. PC: polarization controller, PBS: polarization beam-splitter,
    AM: amplitude modulator, AOM: acousto-optic modulator (frequency shifter), EDFA: Erbium-doped fiber amplifier,
    BPF: bandpass filter, OI: optical isolator. Blue lines: optical fibers, Green lines: polarization-maintaining (PM)
    optical fibers, Black lines: RF connections. The actual resonator is highlighted with a green
    background. Vertical (orange) and horizontal (blue) double-arrows symbolize the two different polarization
    modes at various places in the setup.}
  \label{fig:setup}
\end{figure*}

\noindent Figure~\ref{fig:setup} illustrates the experimental setup. The passive ring resonator (highlighted in
green) has a total length of about $10.5$~m, corresponding to a free spectral range (FSR) of $19.76$~MHz ($\pm
20$~kHz) and a round trip time $t_\mathrm{R}$ of $50.60$~ns ($\pm 50$~ps). It is built around a fiber coupler
(beam-splitter) that recirculates 90\,\% of the intracavity light, and allows for the injection of the coherent
driving field (entering from the top right in the figure). Another 1\,\% tap coupler extracts a small fraction of the
intracavity field for analysis through three photodiodes monitoring respectively the total output intensity as well
as the individual intensities of the two polarization modes (more details are given below). The rest of the resonator
is made up of ``spun'' single-mode silica optical fiber, a type of fiber which presents very little polarization
anisotropy (birefringence) \cite{barlow_birefringence_1981}. At the 1550~nm driving wavelength, that fiber exhibits
normal group-velocity dispersion ($-40\ \mathrm{ps/nm/km}$), which has been selected to avoid modulation
instabilities \cite{agrawal_nonlinear_2013}. The measured resonator finesse is $24.1$ ($\pm 0.1$), amounting to total
losses of 26\,\% per roundtrip. The associated photon lifetime and resonance linewidth are, respectively, about
$4\,t_\mathrm{R}$ and 820~kHz. The resonance linewidth is much broader than that of our driving laser, a
distributed-feedback cw Erbium-doped fiber laser (Koheras AdjustiK E15), with a linewidth ${< 1}$~kHz, therefore
guaranteeing coherent driving. The driving laser frequency can be tuned via a piezo-electric actuator, which offers a
simple way to scan the cavity resonances.

Due to unavoidable fiber bending and other imperfections, the two polarization modes of the resonator are slightly
linearly coupled \cite{pierce_coupling_1954}. As a consequence, the interactions between the two modes are not only
dependent on the modal intensities as described above, but are also phase sensitive \cite{cao_experimental_2017}. To
avoid this complication, we drive the two polarization modes with slightly different carrier frequencies. At the same
time, we purposefully introduce some fixed birefringence in the resonator through an intracavity polarization
controller \cite{lefevre_single-mode_1980}, $\mathrm{PC_{intra}}$ in Fig.~\ref{fig:setup}, to counterbalance the
associated difference in wavenumbers, and to realize effective isotropic (or close to isotropic) conditions for the
two driven polarization components. In practice, we have separated the two families of orthogonally polarized cavity
resonances by about 45\,\% of the FSR; the precise value is not critical. The dual carrier driving field is prepared,
ahead of injection into the resonator, by splitting the output of the single frequency driving laser into two
components with a polarization beam-splitter (PBS), and frequency-shifting one of these with an acousto-optic
modulator (AOM). The other path incorporates a variable attenuator (V) set to introduce the same losses as the AOM.
The two components are then recombined in a second PBS. Together with the use of polarization-maintaining fibers
(shown in green in Fig.~\ref{fig:setup}) in between the two PBSs, no polarization-dependent losses are introduced.
Behind the second PBS, the two carriers have orthogonal polarization and are mapped onto the two resonator modes
using another PC (labeled $\mathrm{PC_{map}}$ in Fig.~\ref{fig:setup}), so that each drive a separate mode.
$\mathrm{PC_{map}}$ is adjusted with the AOM off, which suppresses one of the polarization components of the driving
beam, so as to excite a single family of cavity resonances, carefully canceling any trace of the other (orthogonally
polarized) family through observation of the total output intensity.

Our experimental arrangement enables simple and reproducible control of the two asymmetry parameters. On the one
hand, adjusting the RF frequency applied onto the AOM (around 80~MHz) controls the effective isotropy, specifically
the difference in wavenumbers $\delta\Delta$ with which the two polarization components propagate inside the
resonator. On the other hand, controlling the polarization state of the beam ahead of the first PBS changes the ratio
of driving power between the two modes, or equivalently the driving ellipticity~$\chi$, without affecting the total
driving power. This is achieved with an electronic polarization modulator, complemented with a manual bias
($\mathrm{PC}_\chi$).

To calibrate the measurement of $\delta\Delta$, we first observe the AOM frequency at which the linear resonances of
the two polarization modes overlap and have maximal total peak intensity; this point corresponds to $\delta\Delta=0$.
Note that this calibration stage is performed when the resonator is operated purely in the linear regime. From that
starting position, any change $\Delta f$ in the frequency applied to the AOM corresponds to a change of
$\delta\Delta$ of $2\mathcal{F} (\Delta f/\mathrm{FSR})$ (see Appendix~A). With the uncertainties quoted above for
$\mathcal{F}$ and $\mathrm{FSR}$, this is obtained to within $0.5$\,\%.

Because of environmental fluctuations, the driving ellipticity~$\chi$ typically slowly varies over time at the input
of the resonator. To stabilize the system against these perturbations, we measure and monitor~$\chi$ close to the
resonator input, and apply appropriate feedback to the polarization modulator through a PID controller. The value of
$\chi$ is obtained by tapping 1\,\% of the driving beam, and by measuring the intensities of the two polarization
components of the light, separated with another PBS, with two carefully calibrated photodetectors. A PC
($\mathrm{PC_{monitor}}$ in Fig.~\ref{fig:setup}) placed ahead of the PBS is adjusted such that each diode is only
sensitive to one particular cavity polarization mode. Specifically, this is obtained by verifying that one of the
photodiode reads zero when the AOM is off (a similar procedure was used to separate the modes at the output). We have
made sure that the photodetectors are operated strictly in a linear regime. Also, we have measured a calibration
factor that corrects for the small difference in responsitivities between the two diodes, so that we get the same
reading when they are illuminated by the same intensity. Finally, before each set of measurements, we carefully
measure the zero level of both diodes. $\tan^2\chi$ is then obtained as the ratio of the two photodiodes readings
after zeroing and responsitivity correction. This leads to the value of $\chi$ with an uncertainty that we estimate
at less than $0.5^\circ$. The experimental setup incorporates a second feedback loop (not shown in
Fig.~\ref{fig:setup}) that offers the possibility to lock the driving laser at a set detuning from resonance, using
the method of Ref.~\cite{nielsen_invited_2018}. Note that changing the laser frequency changes the two detunings
$\Delta_+$ and $\Delta_-$, or equivalently the wavenumbers of the two driven polarization components, by the same
amount and does not introduce any asymmetry. When both feedback loops are engaged, all the parameters of the
resonator are quantifiably controlled and stable.

Finally, to reach more easily the peak power level necessary to observe SSB, the resonator is synchronously driven by
flat-top $1.04$~ns long pulses carved into the cw beam of our driving laser with an amplitude modulator (AM). For
reasons explained below, two such pulses are launched per roundtrip, separated by $24.5$~ns. The separation is chosen
large enough to minimize unwanted ripples in the AM driving electronics, while at the same time avoiding the acoustic
echo generated down the optical fibers by the leading pulse and that would affect the shape of the trailing pulse for
separations in the 20--22~ns range \cite{jang_ultraweak_2013}. The use of pulses also avoids the detrimental effect
of stimulated Brillouin scattering that is typical of silica optical fibers, and which would otherwise deplete the
driving beam \cite{agrawal_nonlinear_2013}. Calibration of the normalized driving power~$X$ was obtained by observing
the nonlinear shift of the resonance as a function of driving power. As our results do not fundamentally depend on
$X$ (as long as it is set above the SSB threshold), for simplicity it was kept at the same level for all the
measurements presented below. Specifically, we used $X=10.8$, which corresponds to about $2.7$~W of peak power
(equivalent to 110~mW of total averaged power) at the resonator input. The nonlinear cross-coupling coefficient $B$
was obtained from the ratio of the nonlinear shift of the cavity resonance peak for balanced driving conditions
($\chi=45^\circ$) to that observed when only one mode was driven ($\chi = 0^\circ$). That ratio is $(1+B)/2$, and is
independent of driving power. Two separate measurements gave values of $B$ of $1.55$ and $1.6$, and the value was
refined to $B=1.57$ ($\pm 0.01$) by fitting experimental data to the numerical model. Note that the occurrence of
asymmetric balance does not depend critically on the actual value of~$B$.

\section*{Results}

\begin{figure}
  \centerline{\includegraphics[width=\columnwidth]{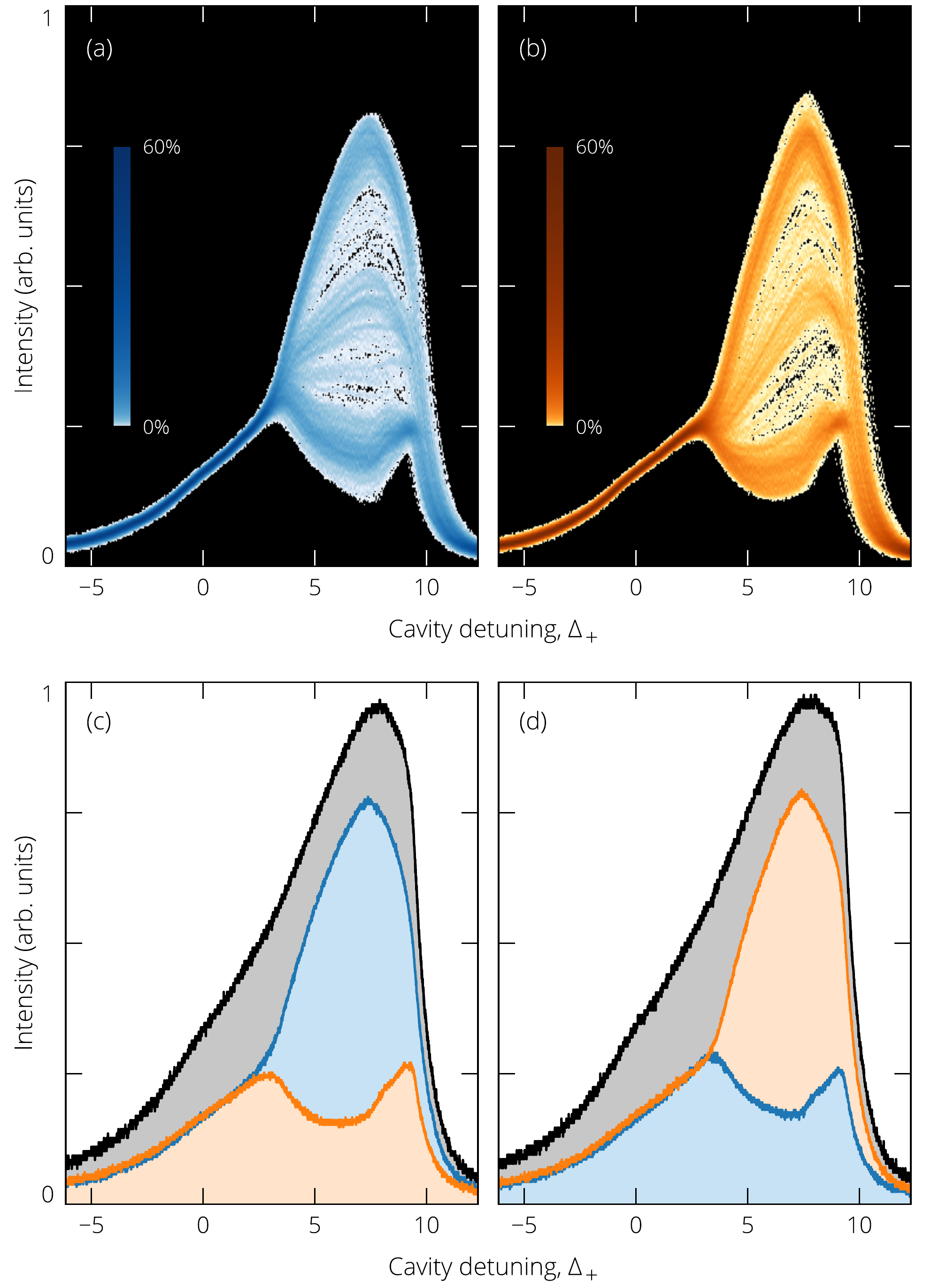}}
  \caption{Nonlinear resonances in the spontaneous symmetry breaking regime illustrated by plotting
    averaged modal intensities (blue and orange curves) recorded as the cavity detuning is ramped up in symmetric
    conditions ($\chi=45^\circ$, $\delta\Delta=0$). In (a) and~(b), measurements are taken over several
    hundreds of scans, and the data are presented as color-coded \%-age-of-occurrence histograms.
    (c),(d) Two individual scans highlighting the anti-correlated behavior of the modal intensities,
    selected to illustrate the mirror symmetry of the system. The total intensity is also shown as black
    curves. Note: the orange curves are associated with the component up-shifted by the AOM.}
  \label{fig:bubble}
\end{figure}

\noindent We start by characterizing our system in symmetric conditions: the driving ellipticity~$\chi$ is maintained
at $45^\circ$ by the feedback loop while $\delta\Delta$ is set to zero. To this end, we repeatedly scan the driving
laser frequency across several cavity resonances while resolving the two polarization modes (Fig.~\ref{fig:bubble},
blue and orange curves respectively). Here, the modal intensities are measured with slow photodiodes that do not
resolve the individual driving pulses. Note that similar results would be obtained by scanning the driving power
\cite{haelterman_polarization_1994}. In Figs.~\ref{fig:bubble}(a) and~\ref{fig:bubble}(b), we present histograms of
each mode intensity obtained by combining seven measurements comprising about 100 resonance scans each. Remarkably,
while the two modes have equal intensities near the base of the resonance (in line with the symmetric conditions),
the peak of the resonance exhibits a high degree of variability \cite{juel_effect_1997}. We observe that high
intensity in one mode always correlates with low intensity in the other mode, as is made evident by the two
individual scans shown in Figs.~\ref{fig:bubble}(c),(d): symmetry is markedly broken. Note that the scanning rate of
1~FSR per millisecond, corresponding to one resonance linewidth per 200~photon lifetime, is slow enough to guarantee
that transients do not affect the mode selection.

In Figs.~\ref{fig:bubble}(c),(d), we have also plotted the total output intensity (black curves) measured at the
resonator output. The total intensity does not display any sign of the underlying pitchfork instability, thus
highlighting that the SSB studied here is a purely polarization phenomenon. To the best of our knowledge, this is the
first observation of a polarization SSB in passive driven resonators, which was theoretically predicted more than
three decades ago \cite{areshev_polarization_1983, haelterman_polarization_1994}. Additionally, a comparison between
Figs.~\ref{fig:bubble}(c) and~\ref{fig:bubble}(d) highlights the very high degree of mirror symmetry in our system,
with the two sets of curves very nearly matching each other. The high variability in Figs.~\ref{fig:bubble}(a,b) can
be interpreted as due to different subparts of the two pulses circling the resonator spontaneously breaking their
symmetry one way or the other, randomly. Since these are not resolved by our slow photodetectors, this leads to
\emph{averaged} intensities spanning the entire range of values between those observed when pulses switch as a whole
[panels (c) and~(d) correspond to that latter case] even though the system has only two stationary solutions that are
mirror of each other \cite{haelterman_polarization_1994}. This latter fact will be formally confirmed below.

Departing from symmetric conditions through a change in $\chi$ or $\delta\Delta$ leads to the disappearance of the
behavior reported in Fig.~\ref{fig:bubble}. The system then always favors the same mode: resonance scans look either
like the one presented in Fig.~\ref{fig:bubble}(c), or the one in Fig.~\ref{fig:bubble}(d), depending on the
direction of the change \cite{benjamin_bifurcation_1978, golubitsky_analysis_1979, *golubitsky_singularities_1985}. A
secondary state, an almost mirror image of the first one, is never excited spontaneously although it might be present
in the system \cite{ball_understanding_2001, gelens_multistable_2010}. In order to probe its existence under
asymmetric conditions, and to identify all the cw stationary solutions of the system for each set of asymmetry
parameters ($\chi$~and~$\delta\Delta$), we proceeded as follows. With the detuning locked and total driving power
kept constant, we measured the instantaneous output intensities of the two polarization components with
10~GHz-bandwidth photodiodes that resolve individual pulses, and acquired the data over about 8000 successive cavity
round-trips with a~40~Gsample/s oscilloscope. In the middle of these acquisitions, we apply strong perturbations to
the two driving pulses through the polarization modulator (see top left of Fig.~\ref{fig:setup}) for about
100~roundtrips. The two pulses driving the resonator are subject to opposite perturbations to maximize the chance
that one of the pulses will switch to the other solution, irrespective of which solution is initially spontaneously
excited. For each oscilloscope trace, the instantaneous intensity levels of all the recorded pulses are then built
into histograms, separately for the two pulses driving the resonator, and before and after the perturbation. Care is
taken to avoid any transients following the perturbation, and pulses that are only partly switched. The maxima of the
histograms allow us to identify the intensity levels of the steady-state stable asymmetric solutions for each
polarization. Note that the use of two driving pulses provides two independent realizations of the experiment in the
same conditions and clearly reveals the coexistence of the identified states. These measurements are repeated as we
step $\chi$ (by adjusting the corresponding feedback loop setting point) and for different values of $\delta\Delta$.

\begin{figure}
  \centerline{\includegraphics[width=\columnwidth]{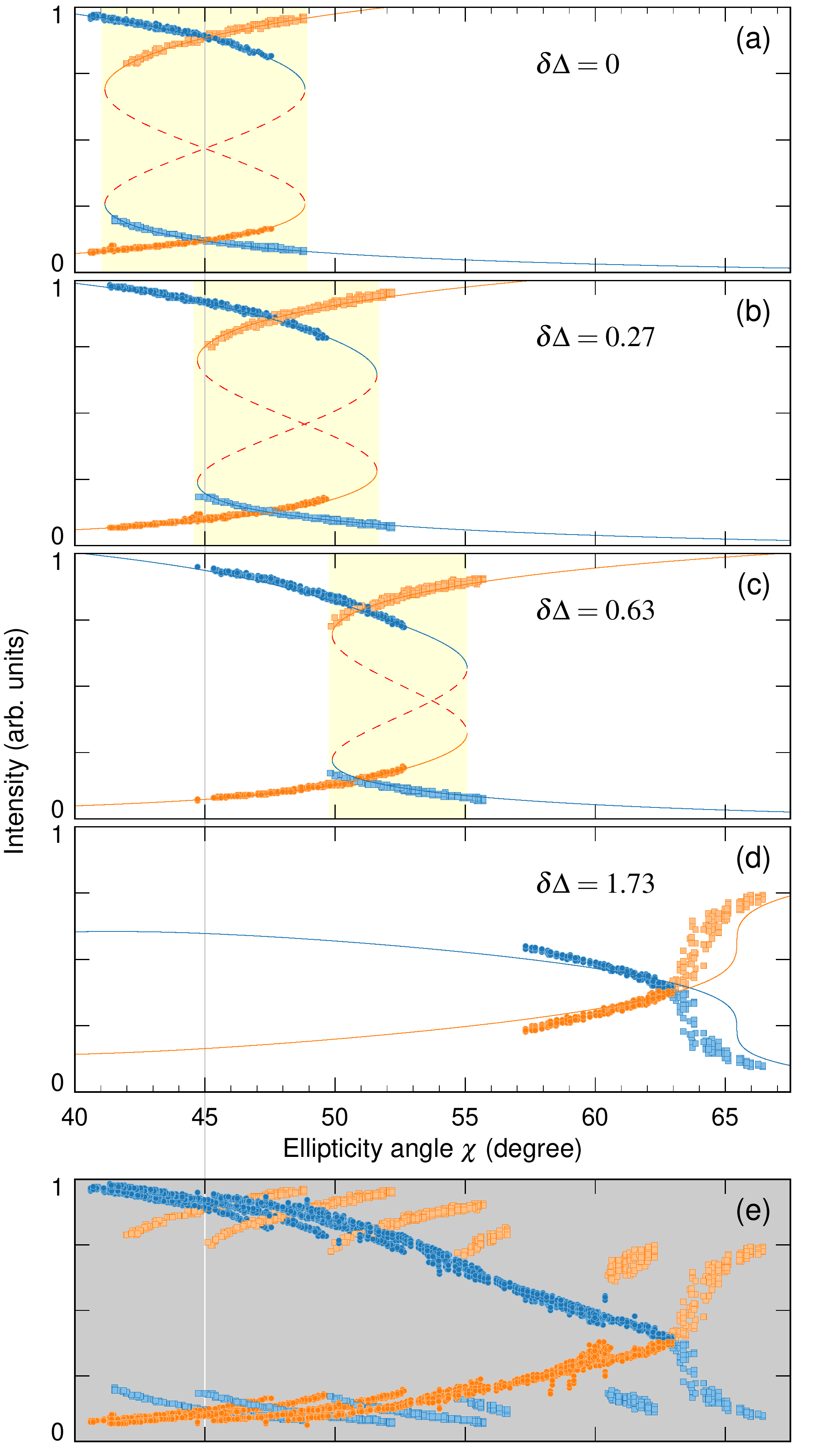}}
  \caption{Measured modal intensities of identified stationary SSB solutions versus driving ellipticity~$\chi$ for
    different values of $\delta\Delta$ and fixed frequency detuning~$\Delta_+=5.45$. (a)--(d) Each solution is
    associated with two points (one for each mode, blue and orange) while different solutions are distinguished by
    using respectively darker round (when the ``blue'' mode dominates) and lighter square (``orange'' dominates)
    symbols. Smooth lines correspond to theoretical predictions, with dashed lines denoting unstable states. The yellow
    band highlights the region of co-existence of states with opposite symmetries. (e) Combination of all the
    experimental results shown in (a)--(d), plus extra data obtained for $\delta\Delta = 1.00$ and $1.49$.}
  \label{fig:resu}
\end{figure}

The results are summarized in Fig.~\ref{fig:resu}, where we plot the modal intensities of the identified stationary
solutions (rounds and squares). The error level is indicated by the size of the markers. Panel~\ref{fig:resu}(a) has
been obtained with $\delta\Delta \simeq 0$. The two solutions identified for $\chi=45^\circ$ are exact mirror images
of each other, as expected from perfect symmetry conditions (see Fig.~\ref{fig:bubble}). These measurements also
confirm that in these conditions the system presents two and only two stable asymmetric solutions, validating our
interpretation of the histograms above. As $\chi$ is varied around that point, we can observe that both solutions
continue to exist, even though their degeneracy is lifted. Theoretical predictions (smooth curves) obtained by
looking for the stationary solutions of Eqs.~(\ref{eq:LL1})--(\ref{eq:LL2}) for the experimental parameters agree
very well with the measurements ($X=10.8$, $\Delta_+=5.45$, and $B$ fitted to $1.57$). Note that some solutions
predicted theoretically are not observed in the experiment because they are only metastable in our pulsed driving
conditions \cite{garbin_experimental_2017}. The range of coexistence is highlighted as a yellow band, and is
reasonably wide, covering almost $10^\circ$ of ellipticity change. Outside that band, however, the driving asymmetry
becomes too strong, and only one solution remains: the symmetry breaking instability effectively disappears. In
Fig.~\ref{fig:resu}(b), we have introduced some asymmetry between the wavenumbers. Interestingly, we observe that the
coexistence region seems to simply shift to a different range of values of $\chi$. In particular, there exists a
value of $\chi \neq 45^\circ$ where the two solutions again appear to be mirror images of each other (where the blue
and orange curves intersect). The two asymmetries, in $\chi$ and $\delta\Delta$, are now balancing each other. The
balance can be realized continuously, i.e., for every value of $\delta\Delta$ within a certain range (see Appendix~B
for a geometric interpretation). It is also very robust: in Fig.~\ref{fig:resu}(c), $\delta\Delta$ is large enough
for the coexistence region not to even overlap with the balanced driving condition at $\chi=45^\circ$. Eventually,
when too much asymmetry is present [Fig.~\ref{fig:resu}(d)], the coexistence region disappears: symmetry breaking is
well and truly destroyed and cannot be restored through a balance of asymmetries. Figure~\ref{fig:resu}(e) highlights
how all the experimental data in Figs.~\ref{fig:resu}(a)--(d) (plus some extra measurements) fit together. We must
stress that the theoretical fits shown in Fig.~\ref{fig:resu} have all been obtained for the same parameters values
and with the values of $\delta\Delta$ directly measured experimentally. This makes the overall agreement quite
remarkable.

\begin{figure}[b]
  \centerline{\includegraphics[width=\columnwidth]{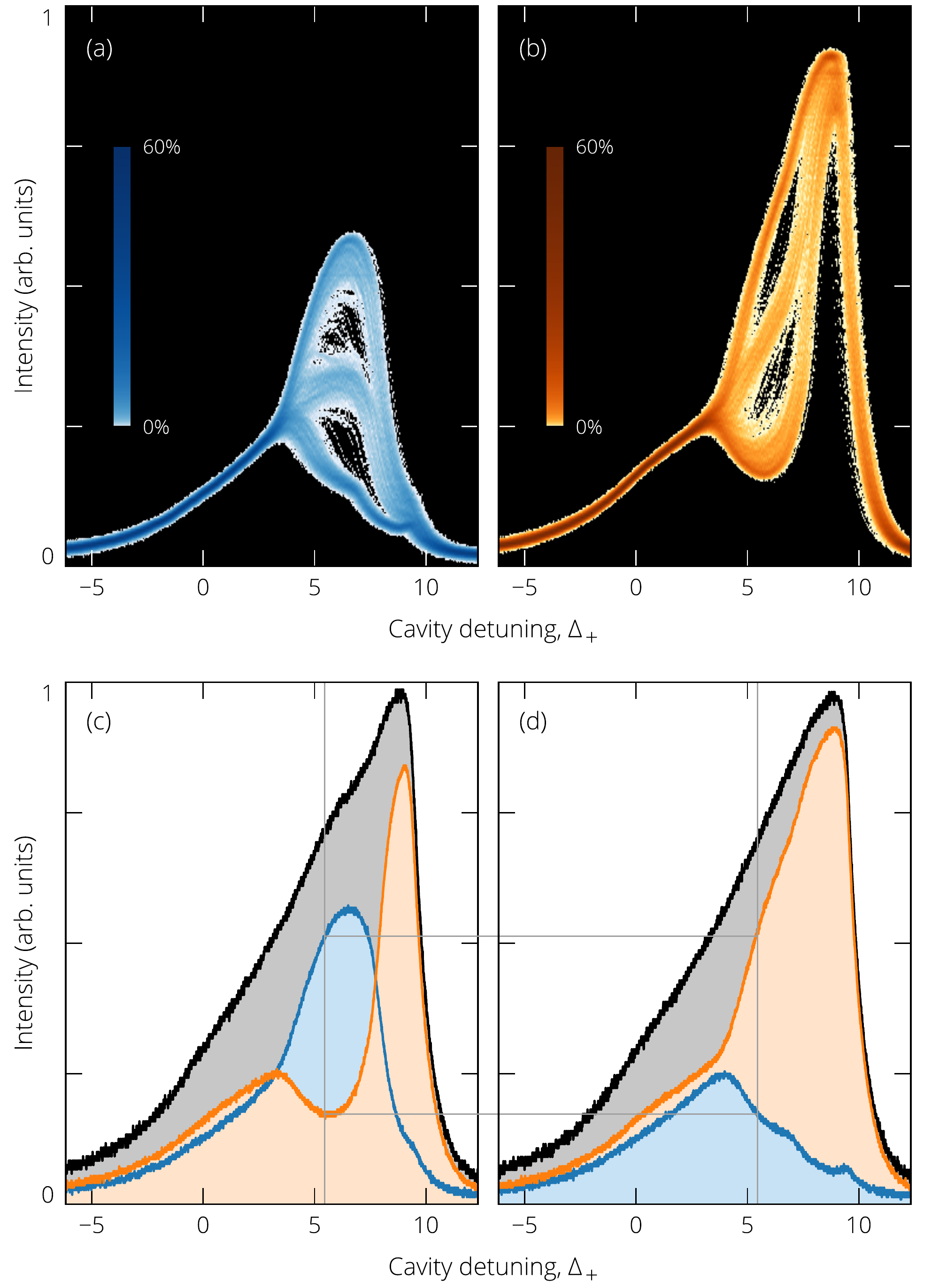}}
  \caption{Same as Fig.~\ref{fig:bubble} but under conditions of asymmetric balance, with $\delta\Delta=0.63$
    [same value as in Fig.~\ref{fig:resu}(c)] and $\chi = 53.5^\circ$. These measurements illustrate that an SSB-like
    response can be found under asymmetric conditions when asymmetries in $\chi$ and $\delta\Delta$ are critically
    balanced. In particular, a ``blue'' dominated solution can still be spontaneously excited, (c), even though the
    ``orange'' mode is driven stronger. Grey lines in (c), (d) highlight how the two solutions observed for
    $\Delta=5.45$ are close mirror of each other, corresponding to the crossing point in Fig.~\ref{fig:resu}(c).}
  \label{fig:bubble-asym}
\end{figure}

\begin{figure*}
  \centerline{\includegraphics[width=\textwidth]{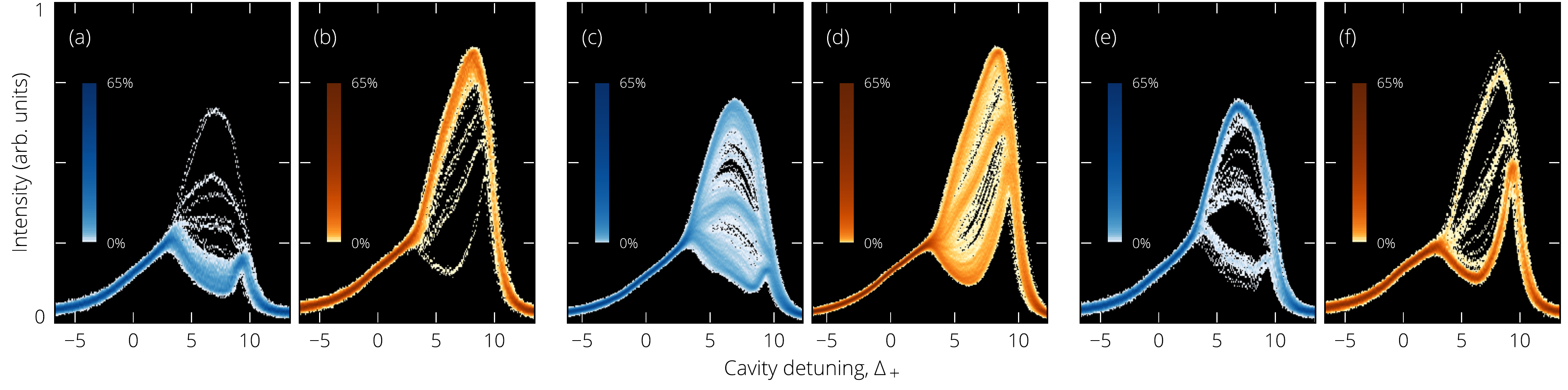}}
  \caption{Resonance scan histograms of average modal intensity levels (same color scheme as other figures) observed
    when bracketing the asymmetric balance condition found with $\chi=48.5^\circ$ (1:$1.3$ driving power ratio).
    (a,b) $\delta\Delta = 0.27 - 0.015$, and an ``orange'' dominated solution is favored.
    (c,d) $\delta\Delta = 0.27$, asymmetric balance is realized.
    (e,f) $\delta\Delta = 0.27 + 0.015$, and the solution is dominated by the ``blue'' mode. The
    histograms in balanced conditions include data from $650$ forward resonance scans each, while the other includes
    about 110 scans each. A change in $\delta\Delta$ of $0.015$ corresponds to a change of 6~kHz in the AOM driving
    frequency.}
  \label{fig:bubble-asym-plusminus}
\end{figure*}

To explore further the regime of asymmetric balance, we performed additional resonance scan measurements with
parameters close to those where we find mirror-like solutions. For $\delta\Delta=0.63$ [corresponding to
Fig.~\ref{fig:resu}(c)], this is illustrated in Fig.~\ref{fig:bubble-asym}, using the same format as in
Fig.~\ref{fig:bubble}. Remarkably, the histograms of Figs.~\ref{fig:bubble-asym}(a,b) reveal that, for a critical
value of driving ellipticity $\chi \simeq 53.5^\circ$, and despite the strong asymmetries, the system presents again
a high degree of variability similar to that observed under symmetric conditions. The two individual scans plotted in
Figs.~\ref{fig:bubble-asym}(c) and~\ref{fig:bubble-asym}(d) highlight that the observed variability stems from the
random selection of one of two solutions of opposite symmetries, i.e., in which a different polarization mode
dominates. The fact that the ``orange'' mode is driven more strongly than the ``blue'' mode (the driving ellipticity
corresponds here to a factor of about $1.8$ difference in driving intensity) is evident in
Fig.~\ref{fig:bubble-asym}, yet it does not preclude a ``blue'' dominant solution to be \emph{spontaneously} excited
[Fig.~\ref{fig:bubble-asym}(c)]. Similar to the symmetric case, and perhaps more remarkably, we again observe no sign
of the instability in the total intensity [black curves in Figs.~\ref{fig:bubble-asym}(c),(d)]. These results show
that asymmetric balance means more than just restoring mirror-like solutions. A pitchfork-like spontaneous-selection
dynamics is recovered when asymmetries are balanced. We note that this behavior agrees with what would be expected
from the two-parameter unfolding of an imperfect pitchfork bifurcation \cite{golubitsky_analysis_1979,
*golubitsky_singularities_1985}. The bifurcation point in Figs.~\ref{fig:bubble-asym}(a,b) where the mode selection
takes place corresponds to what has been referred to as the ``transcritical point'' in
Refs.~\cite{benjamin_bifurcation_1978, *benjamin_bifurcation_1978-1}. Note that the critical value of $\chi$ found in
Fig.~\ref{fig:bubble-asym} ($53.5^\circ$) is slightly different with that observed to give mirror-like solutions in
Fig.~\ref{fig:resu}(c) ($51^\circ$), but this is consistent with the dependence of the critical point on the cavity
detuning~$\Delta_+$ and matches numerical predictions.

When slightly offset from asymmetric balance conditions, the system of course always favors one of the two solutions
and, perhaps not unsurprisingly, a different mode is found to dominate for opposite directions of change. We
illustrate this point in Fig.~\ref{fig:bubble-asym-plusminus} for a different value of $\chi=48.5^\circ$, for which
asymmetric balance is realized with $\delta\Delta=0.27$ [corresponding to Fig.~\ref{fig:resu}(b)].
Figures~\ref{fig:bubble-asym-plusminus}(c,d) show histograms in balanced conditions for these parameters; the results
are similar to those shown for $\delta\Delta = 0.63$ in Fig.~\ref{fig:bubble-asym}. In
Figs.~\ref{fig:bubble-asym-plusminus}(a,b), $\delta\Delta = 0.27 - 0.015$, and the system preferentially selects the
solution for which the ``orange'' mode dominates, while the opposite occurs when $\delta\Delta = 0.27 + 0.015$
[Figs.~\ref{fig:bubble-asym-plusminus}(e,f)]. The non-zero probability of occurrence of the other solution is due to
noise in the system, and to the proximity to the critical point. We again note that all these findings align with
what would be expected from a standard pitchfork, except that we are observing these behaviors in the presence of
asymmetries.

\begin{figure}
  \centerline{\includegraphics[width=\columnwidth]{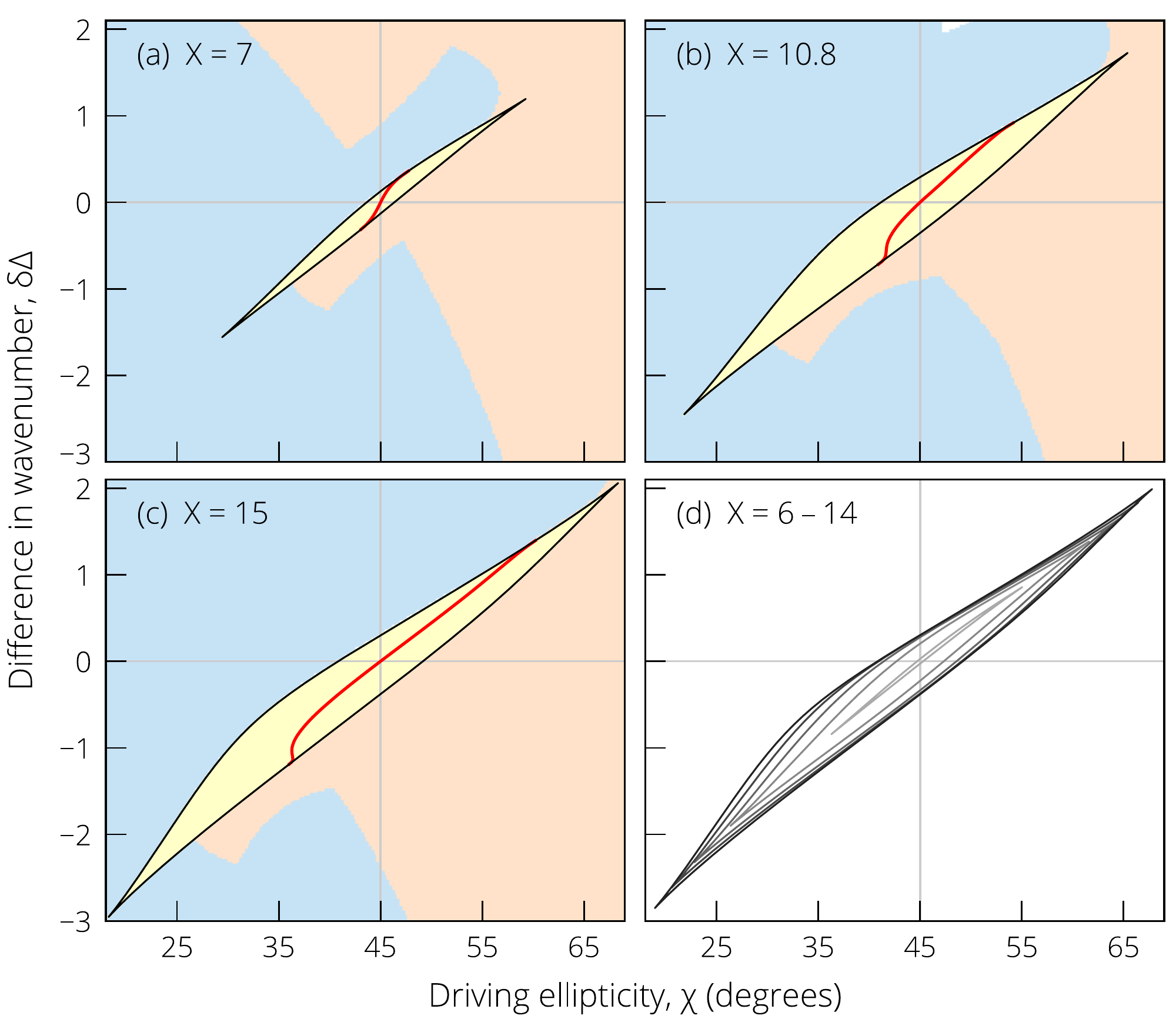}}
  \caption{(a)--(c) Region of co-existence (yellow, surrounded by black line) of symmetry-broken mirror-like
    solutions in the $(\chi,\delta\Delta)$ parameter space evaluated numerically for different driving powers~$X$,
    with $\Delta_+ = 5.45$ and $B=1.57$. The condition for asymmetric balance is shown as a red curve in each graph.
    Blue- and orange-shaded areas indicate which mode dominates outside the co-existence region. The white area
    indicates a region where there are no stable homogenous solutions. (d) Superposition of the limit of the
    co-existence regions, for driving power $X=6$--14, in step of~2 (outward).}
  \label{fig:range}
\end{figure}

At this point, it must be clear that our findings are not specific to the parameters of Figs.~\ref{fig:bubble-asym}
or~\ref{fig:bubble-asym-plusminus}: spontaneous selection and a pitchfork-like dynamics can be restored over a
continuous range of values of $\delta\Delta$ or driving power around those illustrated. In fact, no parameters need
to take a specific value for the effects described here to be observed. This makes clear that our observations are
not the result of an accidental symmetry. To illustrate this point further, we have calculated (using numerical
continuation~\cite{auto}) the limits of the regions of the $(\chi,\delta\Delta)$ asymmetry parameter space where
Eqs.~(\ref{eq:LL1})--(\ref{eq:LL2}) exhibit co-existence of two stable mirror-like symmetry broken solutions, as well
as the conditions in which asymmetric balance can be achieved. This is illustrated in Figs.~\ref{fig:range}(a)--(c),
for different values of driving power~$X$ and for a fixed detuning of the $+$ mode, $\Delta_+=5.45$, matching the
experimental conditions. The black curves delineate the co-existence region, also highlighted with a yellow
background; this region corresponds to that highlighted in a similar way in Fig.~\ref{fig:resu}. In the co-existence
region, we have also plotted as a red curve the combination of parameters for which asymmetric balance is achieved
(strictly speaking, where the upper intensities of the two co-existing solutions match; see also Appendix~B).
Overall, these plots illustrate the wide range of parameters for which asymmetric balance can be realized. Just
outside the co-existence region, only one stable symmetry broken solution survives; the background color in
Figs.~\ref{fig:range}(a)--(c) then indicates which polarization mode of the intracavity field is the most intense
(blue when $|E_+|^2>|E_-|^2$, and orange otherwise). For large asymmetries, far from the co-existence region, complex
bistable cycles can occur, leading again to multiple co-existing stable solutions, but we find that these solutions
never have a mirror-like association: they all exhibit the same dominant mode (we exclude the lower state in this
analysis), and we use the same color scheme. We note that the plots in Figs.~\ref{fig:range}(a)--(c) are not
symmetric with respect to the transformation $(\delta\Delta \rightarrow -\delta\Delta, \chi \rightarrow
90^\circ-\chi)$ but this is simply a result of maintaining $\Delta_+$ constant. Symmetric diagrams would be found if
instead the \emph{average} detuning, $(\Delta_+ + \Delta_-)/2$, was to be kept constant. Finally,
Fig.~\ref{fig:range}(d) is a superposition of the co-existence ranges observed for different driving powers, and show
how that range widens with increasing driving power.

\section*{Conclusion}

\noindent In summary, we have studied experimentally a system presenting an SSB instability in presence of two
asymmetry parameters. By systematically tracking the different stationary solutions of the system with controlled and
quantified asymmetries, we have observed that, while the SSB dynamics is destroyed by one asymmetry, it can in
practice be restored by a second, properly balanced asymmetry. To the best of our knowledge, this is the first
experimental realization of a restoration of an SSB-like dynamics through a controlled balance of two asymmetries. We
note that the results presented here are restricted to the $\mathrm{Z}_2$ symmetry group, and that more work is
needed to generalize them to other symmetry groups. However, given the importance and ubiquity of SSB in the physical
sciences, our work is still relevant to numerous fields. In particular, it could be extended to other multimode
systems, and it shows that applications of SSB in sensing based on real, necessarily imperfect, physical platforms,
could potentially use asymmetry as an additional degree of freedom to reach divergent sensitivity
\cite{kaplan_enhancement_1981, woodley_universal_2018}. This overlaps with theoretical investigations showing that
so-called exceptional points, recently heralded at providing greatly enhanced sensitivity in optical sensors
\cite{hodaei_enhanced_2017, *chen_exceptional_2017, lai_observation_2019}, can be found under generic asymmetric
conditions \cite{kominis_asymmetric_2016, *kominis_exceptional_2018}. More generally, other studies have also shown
that asymmetry is sometimes necessary for behavioral symmetry \cite{nishikawa_symmetric_2016, majhi_asymmetry_2018,
hart_topological_2019}. We must also point out that our experiment, performed in the context of nonlinear optics,
constitutes the first observation of the polarization symmetry breaking of a passive, coherently driven nonlinear
resonator \cite{areshev_polarization_1983, haelterman_polarization_1994}. It paves the way to the robust realization
of persistent polarization domain walls, which could be applied to novel computing schemes
\cite{gilles_polarization_2017}. Finally, we note that, because optical fiber ring resonators are formally equivalent
to Kerr microresonators, our implementation is amenable to miniaturization and integration
\cite{cao_experimental_2017, del_bino_symmetry_2017}.

\bigskip

\begin{acknowledgments}
  \noindent We thank Alexander Nielsen for help with the last histogram measurements, Andrus Giraldo for advice on
  continuation software, and Lewis Hill for useful discussions. We acknowledge financial support from The Royal Society
  of New Zealand, in the form of Marsden Funding (18-UOA-310), as well as James Cook (JCF-UOA1701, for S.C.) and
  Rutherford Discovery (RDF-15-UOA-015, for M.E.) Fellowships. J.F. thanks the Conseil r\'egional de Bourgogne
  Franche-Comt\'e, mobility (2016-9201AAO050S01777).
\end{acknowledgments}

\appendix

\section{Appendix A: Normalization of wavenumbers and cavity detunings}
\label{app:norm}

\noindent The definitions below apply equally to both polarization modes of the resonator ($+$ and $-$) but to
simplify the notations we start by focusing on the $+$ mode. Assuming light driven in that polarization component
propagates with a wavenumber $\beta_+$, that wave accumulates over one roundtrip in the resonator a linear phase
shift $\beta_+ L$ (with respect to the driving field), where $L$ is the resonator length. We define the corresponding
phase detuning $\delta_+ = 2m\pi - \beta_+ L$ as the distance (in phase) to the closest resonance (of index $m$). A
positive value of phase detuning corresponds to a driving beam red shifted with respect to the corresponding linear
resonance. Normalized detuning is defined as $\Delta_+ = \delta_+/\alpha$, where $\alpha$ represents the resonator
losses, specifically half the percentage of power lost per roundtrip. With that notation, the resonator finesse is
simply given by $\mathcal{F} \simeq \pi/\alpha$. The normalized difference in wavenumbers is then defined as
$\delta\Delta = \Delta_+ - \Delta_-$.

\section{Appendix B: Geometric interpretation of asymmetric balance}
\label{app:geom}

\begin{figure*}[t]
  \centerline{\includegraphics[width=\textwidth]{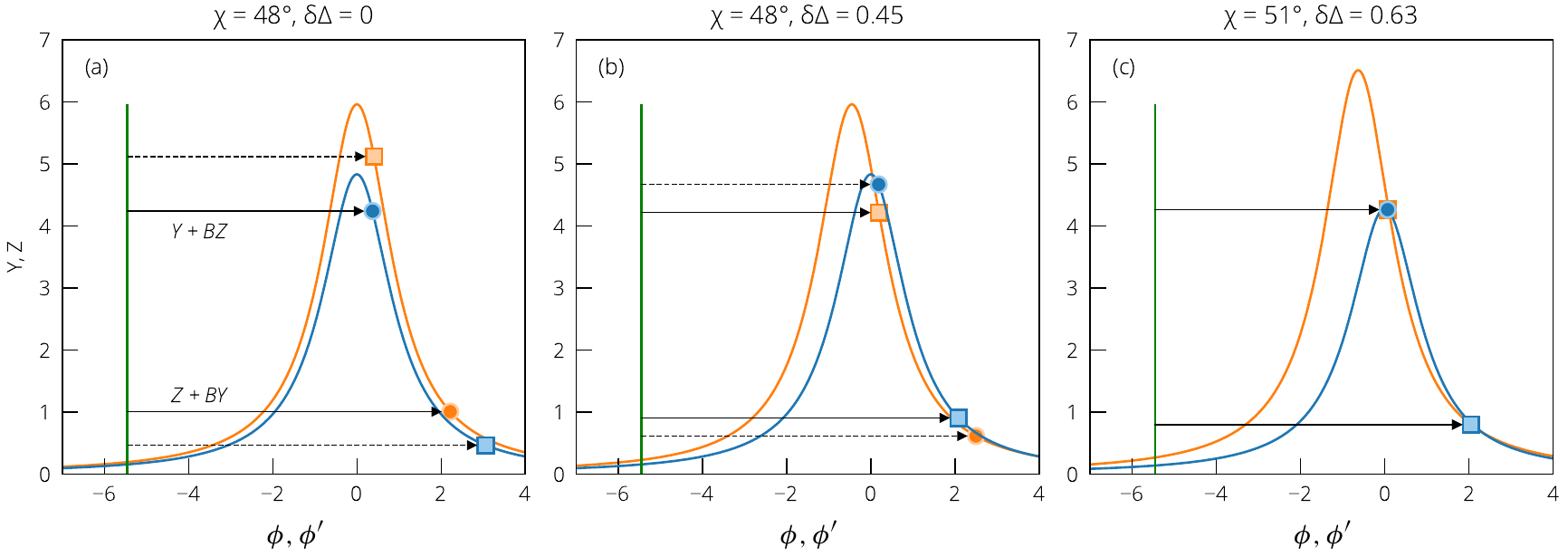}}
  \caption{Graphical interpretation of asymmetric balance. Each panel shows, for different parameter values,
    how the symmetry broken solutions lie with respect to the resonances of the two polarization modes. Each
    solution is associated with two points (one for each mode, blue and orange) while the
    black arrows represent the corresponding (normalized) nonlinear shifts with respect to the driving laser
    frequency (green line at position $\phi=\phi'=-\Delta_+$). Different solutions are distinguished by using
    respectively darker round (when the ``blue'' mode dominates) and lighter square (``orange'' dominates) symbols, as
    well as solid versus dashed arrows for the nonlinear shifts. Asymmetric balance is obtained when the solutions
    coincide with the intersections of the two resonances. Other parameters are: $X=10.8$, $\Delta_+=5.45$, and
    $B=1.57$, as in the experiment.}
  \label{fig:sup:geo}
\end{figure*}

\noindent The conditions in which asymmetric balance is possible can be qualitatively understood by a generalization
to the asymmetric case of the diagram shown in Fig.~\ref{fig:principle}(c), and that explains the origin of
spontaneous symmetry breaking (SSB) in the passive nonlinear Kerr resonator. Starting from
Eqs.~(\ref{eq:LL1})--(\ref{eq:LL2}), we can write the following two equations for the intracavity modal intensity
levels, $Y=|E_+|^2$, $Z=|E_-|^2$, of the stationary ($\partial/\partial t=0$) solutions,
\begin{align}
  Y &= \frac{X \cos^2\chi}{1 + (Y + BZ - \Delta_+)^2}\,,\label{eq:Y}\\
  Z &= \frac{X \sin^2\chi}{1 + (Z + BY - \Delta_+ + \delta\Delta)^2}\,.\label{eq:Z}
\end{align}
These equations are nonlinear and do not have closed form solutions. Their Lorentzian form, in terms of the total
linear and nonlinear phase-shifts, nevertheless illustrate the resonant behaviour of the system. In the \emph{scalar}
case ($\chi=0^\circ$, $Z=0$), there exists a geometric construction of the solution to the remaining non-trivial
equation, which correctly explains all the features of the scalar Kerr bistability
\cite{fraile-pelaez_transmission_1991}. This geometric construction cannot be generalized to the vector case above,
because of the different dependence on $Y$ and $Z$ in the right-hand sides of the two equations. Using some of the
ideas developed in~\cite{fraile-pelaez_transmission_1991}, we can however gain interesting insights about the
stationary states, Eqs.~(\ref{eq:Y})--(\ref{eq:Z}). To this end, we plot the two resonances above, respectively
versus $\phi= Y + BZ - \Delta_+$ and $\phi'= Z + BY - \Delta_+$, along the same axis. This is illustrated for three
examples in Fig.~\ref{fig:sup:geo}. Note how by construction the $Z$ resonance is shifted by~$\delta\Delta$ with
respect to the $Y$ resonance. The difference in amplitude on the other hand reflects the driving ellipticity~$\chi$.

On these diagrams, we have represented the two symmetry-broken solutions calculated numerically for the parameters of
the three examples considered. Each solution $(Y,Z)$ is plotted as a pair of points at coordinates $(\phi,Y)$ (blue)
and $(\phi',Z)$ (orange). By representing the driving laser frequency as a vertical green line at position
$\phi=\phi'=-\Delta_+$, the distances between that line and the different points represent the corresponding
normalized nonlinear phase shifts, $Y+BZ$ (for the ``blue'' mode) and $Z+BY$ (for the ``orange'' mode). The two
solutions are distinguished from each other by using respectively dark round and light square symbols as in
Fig.~\ref{fig:resu}, as well as with solid versus dashed arrows for the nonlinear phase shifts. We can observe that
the symmetry broken solutions always lie on the right side of the resonances \footnote{This is assuming $B>1$.
Symmetry broken solutions of Eqs.~(\ref{eq:Y})--(\ref{eq:Z}) also exist for $B<1$ but these are always unstable, and
are not considered here.}, where the slopes are negative, and opposite from the driving laser (green line). We remark
that, in the case of scalar bistability, that side is associated with the stable upper state. Geometrically, this is
explained by noting that the nonlinear phase shift has to effectively push the driving across the resonance to reach
that state~\cite{fraile-pelaez_transmission_1991}. This correctly ties with the fact that SSB in the passive Kerr
resonator always originates on the upper branch of the bistable response~\cite{haelterman_polarization_1994}.

In the presence of asymmetries, the two symmetry broken solutions are typically not mirror image of each other, and
the points corresponding to the two solutions are distinct. This is in particular the case in presence of a single
asymmetry as in Fig.~\ref{fig:sup:geo}(a), where $\chi>45^\circ$ (the ``orange'' mode is driven stronger than the
``blue'' mode) and $\delta\Delta=0$. Starting from that configuration, and introducing the second asymmetry by
increasing $\delta\Delta$, the two resonances as plotted in our diagram can eventually intersect; see
Fig.~\ref{fig:sup:geo}(b). Asymmetric balance is achieved when one symmetry broken solution matches with these
intersection points [Fig.~\ref{fig:sup:geo}(c)], because at these points $Y=Z$, $\phi=\phi'$, and each solution is
the mirror image of the other. In practice, we find that this match is rarely perfect, but approaching it to within
about~1\,\% in terms of intensities. This occurrence is nevertheless always a telltale sign that the spontaneous
selection between the modes, which is the characteristic of SSB, can be restored for nearby parameters. As stable
symmetry broken solutions always lie on the right side of the resonances, realizing asymmetric balance requires that
the intersection points lie on that same side. Although this geometric argument cannot be turned into a simple
mathematical expression at present, it can still be used to make qualitative predictions. In particular, it shows
that a wavenumber difference $\delta\Delta > 0$ can only be balanced when $\chi>45^\circ$ and vice versa. Also, a
larger absolute value of $\delta\Delta$ requires $\chi$ to depart more significantly from $45^\circ$. Both of these
trends agree with experimental observations.


%

\end{document}